# Distributed Turbo-Like Codes for Multi-User Cooperative Relay Networks

Roua Youssef and Alexandre Graell i Amat
Department of Electronics, Institut TELECOM-TELECOM Bretagne, 29238 Brest, France
Email: roua.youssef@telecom-bretagne.eu,alexandre.graell@telecom-bretagne.eu

*Abstract*— In this paper, a distributed turbo-like coding scheme for wireless networks with relays is proposed. We consider a scenario where multiple sources communicate with a single destination with the help of a relay. The proposed scheme can be regarded as of the *decode-and-forward* type. The relay decodes the information from the sources and it properly combines and re-encodes them to generate some extra redundancy, which is transmitted to the destination. The amount of redundancy generated by the relay can simply be adjusted according to requirements in terms of performance, throughput and/or power. At the destination, decoding of the information of all sources is performed jointly exploiting the redundancy provided by the relay in an iterative fashion. The overall communication network can be viewed as a serially concatenated code. The proposed distributed scheme achieves significant performance gains with respect to the non-cooperation system, even for a very large number of users. Furthermore, it presents a high flexibility in terms of code rate, block length and number of users.

## I. INTRODUCTION

The essential goal of communication systems has always been to enable reliable communication with high data rates. In recent years, cooperative communications for wireless systems has attracted enormous attention. The basic idea of cooperative communications is that all nodes in a wireless network can help each other to transmit information to the destination. Unlike conventional point-to-point communications, this new paradigm offers tremendous advantages such as higher throughput and significant reliability.

The discussion of cooperative communications can be traced back to the 70s when van der Meulen proposed the relay channel [1]. In the past few years, there has been an increasing interest in communication systems with relays. In [2], the authors proposed a novel coding technique for the relay channel called *distributed turbo code*. In the proposed scheme, the source broadcasts to both the relay and the destination. The relay decodes, interleaves and re-encodes the message prior to forwarding. The destination receives two encoded copies of the original message and jointly decodes them by an iterative decoding algorithm. The proposed technique achieves a combined diversity and coding gain. More recently, several works have considered network coding over noisy relay channels. For instance, in [3, 4] joint network-channel coding schemes for the multiple-access relay channel were proposed. In [3], two users transmit to both the destination and the relay, which forwards a combination (a simple XOR) of the received messages from both users.

In [5] an extended scenario where multiple sources transmit to a destination with the help of a relay was considered. Each source encodes its own information using the same block code. The relay decodes the codewords from the sources, places them in a matrix and then re-encodes the columns using another linear block code. The overall scheme can be regarded as a turbo product code. Significant performance gains with respect to the non-cooperation case are achieved. However, the scheme in [5] presents several drawbacks. First, it lacks in flexibility. For instance, changing the number of users requires the use of a different block code at the relay. Also, as for classical turbo product codes, code rate and block length adaptation is difficult.

In this paper, we propose a distributed turbo-like coding scheme for the wireless network with multiple sources considered in [5]. The sources broadcast their information to both the destination and the relay. The relay decodes the received codewords and properly combines and re-encodes them to generate some extra redundancy. The amount of redundancy generated by the relay can be tuned according to performance requirements and/or constraints in terms of throughput and power. The overall communication network can be viewed as a serially concatenated code (SCC), where the outer encoder groups the encoders of the sources and the inner encoder is the relay. Accordingly, the receiver can decode users data using a decoding strategy that resembles the decoding of a SCC. The proposed scheme achieves very low error rates and offers significant performance gains with respect to non-cooperation, even for a very large number of users. Furthermore, it provides a high flexibility in terms of code rate, number of users, throughput and error protection level with respect to the approach in [5].

## II. SYSTEM MODEL

We consider the wireless relay network depicted in Fig. 1. The network consists of $q$ sources, $s_1, \ldots, s_q$, which communicate statistically independent data to a single destination $d$ cooperating through a relay $r$. At source $i$ ($i = 1, \ldots, q$), the information sequence $\mathbf{u}_i$, of length $k_i$ bits, is encoded by encoder $\mathcal{C}_i$ of rate $R_i$ into codeword $\mathbf{x}_i$ of length $n_i = k_i/R_i$ bits, which is transmitted over a wireless channel. Due to the broadcast nature of the wireless channel the relay receives a noisy version of codewords $\mathbf{x}_i$, denoted by $\mathbf{y}_{i,r}$, from the different sources. It then cooperates with them by transmitting its own parity sequence $\mathbf{x}_r$ to the destination. The destination decodes the information of the $q$ sources by jointly exploiting the received sequences $\mathbf{y}_{i,d}$ from the sources and the sequence $\mathbf{y}_{r,d}$ from the relay.

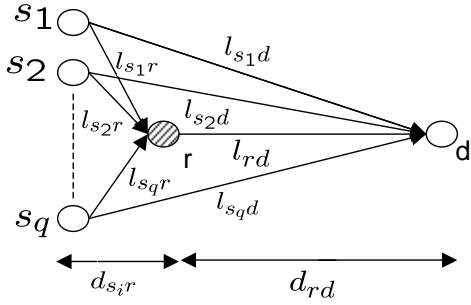

Fig. 1. A wireless relay network: multiple sources transmit to a single destination with the help of a relay.

*A. Channel Model*

As shown in Fig. 1, there are three directed transmission links under consideration: the links from the sources to the destination, $l_{s_i d}$, the links from the sources to the relay, $l_{s_i r}$, and the link from the relay to the destination, $l_{rd}$. We denote by $\gamma_{s_i d}$, $\gamma_{s_i r}$ and $\gamma_{rd}$ the signal-to-noise ratio ($E_b/N_0$) in dB of the $i$-th source-to-destination channel, the $i$-th source-to-relay channel, and the relay-to-destination channel, respectively. Likewise, we denote by $d_{s_i d}$, $d_{s_i r}$ and $d_{rd}$ the distance from source $i$ to the destination, from source $i$ to the relay, and from the relay to the destination, respectively. We also make the following assumptions:

- All sources are at the same distance from the destination, i.e., $d_{s_i d} = d_{sd} \; \forall \; i$. Therefore, $\gamma_{s_i d} = \gamma_{sd}$ for all sources $s_i$.
- The relay is closer to the sources than to the destination. This is a favorable and necessary assumption for decode-and-forward techniques, since it guarantees a low error probability at the relay. Moreover, we assume that $d_{s_i r} = d_{sr}$ for all sources $s_i$, i.e., $\gamma_{s_i r} = \gamma_{sr} \; \forall i$.

The $E_b/N_0$ of the three channels under consideration are linked by

$$\begin{aligned} \gamma_{sr} &= \gamma_{sd} + g_{sr} \\ \gamma_{rd} &= \gamma_{sd} + g_{rd} \end{aligned} \quad (1)$$

where the gains $g_{sr}$ and $g_{rd}$ are due to shorter transmission distances and are given by $g_{sr} = 10n \log(d_{sd}/d_{sr})$ and $g_{rd} = 10n \log(d_{sd}/d_{rd})$, respectively. $n$ denotes the path-loss exponent and it is often assumed to be $2 \leq n \leq 6$ [6,7]. Here, we consider a path-loss exponent of $n = 3.52$ [8].

All channels are assumed to be orthogonal. Orthogonality can be obtained, e.g., via time-division multiple-access (TDMA), i.e., each node transmits in a different time slot. The received signal $y_{i,d}$ at the destination can be expressed as

$$y_{i,d} = h_{i,d} x_i + n_{i,d} \quad (2)$$

where $x_i$ is the transmitted bit from source $s_i$ with energy $E_b$, $h_{i,d}$ is the $i$-th source-to-destination channel coefficient, and $n_{i,d}$ denotes the additive white Gaussian noise. For AWGN channel $h_{i,d} = 1$. On the other hand, if a Rayleigh fading channel is considered, $h_{i,d}$ is a zero-mean complex Gaussian random variable with unit variance. Similar expressions are obtained for the received signals $y_{i,r}$ at the relay and the received signal $y_{r,d}$ at the destination.

## III. ENCODING STRATEGIES AT THE RELAY

The proposed relaying scheme can be regarded to as a *decode-and-forward* scheme. The relay receives a noisy version of the $q$ codewords $\mathbf{x}_i$ from the different sources, it decodes them, and generates an estimate of the transmitted codewords. The estimated codewords $\hat{\mathbf{x}}_i$ are then properly combined, interleaved into $\tilde{\mathbf{x}}$ by an interleaver $\Pi$, and re-encoded by another encoder $\mathcal{C}_e$ prior to transmission. We consider two encoding strategies at the relay.

*A. Strategy A*

The bits of the codeword $\tilde{\mathbf{x}}$ at the output of the interleaver, of length $N = \sum_{i=1}^{q} n_i$ bits, are grouped into groups of $J$ bits and passed to a single parity check (SPC) circuit. This simply forms the sum modulo-2 of the $J$ incoming bits, i.e., for each group of $J$ bits it generates a single bit as their modulo-2 sum. Parameter $J$ can be chosen according to performance requirements, throughput and/or power constraints. The codeword at the output of the SPC, of length $N_{\text{SPC}} = N/J$ bits, is then encoded by a recursive convolutional encoder $\mathcal{C}_e$ (typically of rate $R_e = 1$) and transmitted. The effective code rate of the overall system is $R_{\text{eff}} = K/N'$, where $K = \sum_{i=1}^{q} k_i$, $N' = N + N_r$ and $N_r = N_{\text{SPC}}/R_e$ is the length of the codeword transmitted by the relay. The proposed distributed coding scheme is depicted in Fig. 2.

*B. Strategy B*

The bits of the codeword $\tilde{\mathbf{x}}$ are encoded by a recursive inner encoder $\mathcal{C}_e$ (typically of rate $R_e = 1$) heavily punctured to rate $R_I = R_e/\rho > 1$ through a puncturer and transmitted to the destination. We denote by $\rho$ the permeability ratio of the puncturer, giving the ratio of bits that survive puncturing. Parameter $\rho$ determines the amount of redundancy transmitted by the relay and can be adjusted according to requirements in terms of performance, throughput and/or power. The effective code rate of the overall system is $R_{\text{eff}} = K/N'$, where $N' = N + N_r$ and $N_r = N/R_I$. Note that while the encoder at the relay is of rate higher than one (and therefore non-invertible), the overall distributed coding scheme is of rate $R_{\text{eff}} < 1$, which allows correct recovering of users data. The distributed coding scheme is depicted in Fig. 2.

## IV. THE RELAY NETWORK AS A SERIALLY CONCATENATED CODE

The proposed distributed coding scheme can be seen as a SCC where the outer encoder accounts for the encoders of the $q$ sources, and the inner encoder is the relay. The equivalent representation is depicted in Fig. 3. The information sequences $\mathbf{u}_i$ are concatenated and encoded by the outer encoder $\mathcal{C}_O$. Note that $\mathcal{C}_O$ is time-variant if encoders $\mathcal{C}_i$ are different. The encoded information is transmitted over a channel with $E_b/N_0 = \gamma_{sd}$ (which corresponds to the direct link between the sources and the destination, $l_{sd}$). It is also interleaved and re-encoded by the inner encoder $\mathcal{C}_I$ (which implements strategy A or B). The resulting codeword is transmitted over a channel with $E_b/N_0 = \gamma_{rd}$. Note that the codeword at the input of the inner encoder may contain errors.

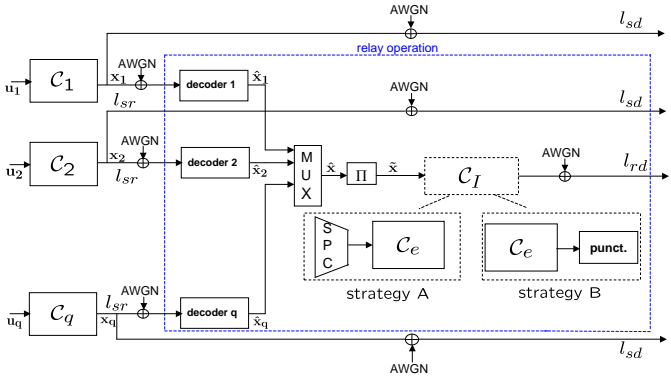

Fig. 2. Block diagram of the proposed distributed turbo-like code. The relay processes information using strategy A or B.

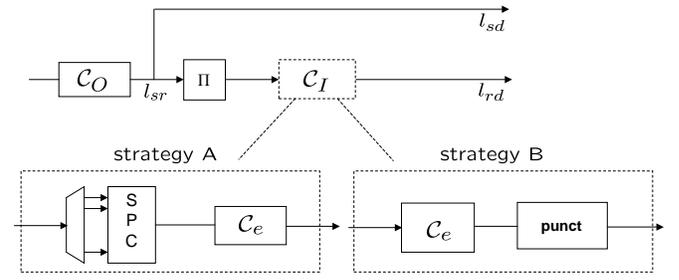

Fig. 3. Equivalent representation of the wireless relay network of figure 2.

From the equivalent representation in Fig. 3 the SCC resulting from strategy A is similar to the coding scheme introduced in [9], nicknamed Flexicode. In [9] a SPC circuit is used before the inner encoder of a SCC to increase the code rate while preserving very good performance in the error floor. Moreover, a copy of the input data is sent directly to the channel, i.e., the code is systematic. Here, the key idea is to use a SPC to group several users and to control the throughput of the relay-to-destination-link, instead of increasing the code rate. The overall scheme is therefore a form of (non-systematic) *distributed* Flexicode. On the other hand, the SCC resulting from strategy B is similar to the SCC scheme proposed in [10]. In [10] a SCC was proposed where the inner encoder was heavily punctured beyond the unitary rate. The proposed scheme allows keeping a very low error floor for all code rates since the interleaver gain is preserved, and achieves much better performance than standard SCCs, especially for high rates. Here, the level of puncturing of the inner code is used to group several users instead of increasing the code rate.

The coding schemes in [9] and [10] are the best known SCCs for high rates. They also allow a high flexibility in terms of code rate. In terms of the wireless network of Fig. 1 *high rates* correspond to a high number of users. Therefore, the proposed schemes are expected to perform very well even for a large number of users.

### A. Decoding of the distributed turbo-like code

According to the equivalent representation in Fig. 3, decoding of users data can be performed using a decoding strategy that resembles the decoding of a SCC: decoding of the information of all sources can be done jointly, exploiting the redundancy provided by the relay in an iterative fashion. One possible (but not unique) scheduling for strategies A and B is as follows:

- The relay-to-destination channel metrics are passed to the inner decoder, and the source-to-destination channel metrics are passed to each source decoder.
- The receiver decodes first the sources and generates extrinsic information on codeword $\hat{\mathbf{x}}$.
- The inner encoder (the concatenation of the SPC and encoder $\mathcal{C}_e$ and the concatenation of encoder $\mathcal{C}_e$ and the puncturer for strategy A and B, respectively) is then decoded by using the extrinsic information on $\hat{\mathbf{x}}$ (properly interleaved) as *a priori* information, and generates extrinsic information on $\tilde{\mathbf{x}}$ which will be used as *a priori* information (after de-interleaving) by the decoders of the sources at the next iteration.
- The process is repeated until the maximum number of iterations is reached or an early stopping criterion is fulfilled.

## V. INFORMATION-THEORETIC LIMITS

In this section, we compute the achievable rates for the proposed scheme, following the approach proposed in [11], where the authors considered the achievable *decode and forward* rate of a 2-users system assuming the time division multiple access relay channel (MARC) model with optimized allocation of the transmission time. In the following, we consider for simplicity the time division MARC with two sources and one relay. We will then give results for the generalized multi-user case.

Let $k_i$, $n_i$, $i = 1, 2$, $N_r$ and $N'$ be as defined in Sections II and III. We define the rate for source $s_i$ as $R'_i = k_i/N'$. We have that $K = k_1 + k_2$, $N' = n_1 + n_2 + N_r$ and the rate of the overall system is $R_{\text{eff}} = R'_1 + R'_2 = K/N'$. The information data of sources $s_1$ and $s_2$ can be decoded reliably at the destination if the following inequalities hold [11]:

$$\begin{aligned} k_1 &\leq n_1 C(\gamma_{s_1 r}) \\ k_2 &\leq n_2 C(\gamma_{s_2 r}) \\ k_1 &\leq n_1 C(\gamma_{s_1 d}) + N_r C(\gamma_{rd}) \\ k_2 &\leq n_2 C(\gamma_{s_2 d}) + N_r C(\gamma_{rd}) \\ k_1 + k_2 &\leq n_1 C(\gamma_{s_1 d}) + n_2 C(\gamma_{s_2 d}) + N_r C(\gamma_{rd}) \end{aligned} \quad (3)$$

where $C$ is the ergodic capacity.

We define the time allocation parameters $\theta_1 = n_1/N'$, $\theta_2 = n_2/N'$ and $\theta_r = N_r/N'$ for sources $s_1$ and $s_2$ and the relay, respectively, with $\theta_1 + \theta_2 + \theta_r = 1$. From (3) the achievable

rate $R_1'$ is given by

$$R_1' = \min\{\theta_1 C(\gamma_{s_1 r}), \theta_2 C(\gamma_{s_2 r})/\sigma,$$
$$\theta_1 C(\gamma_{s_1 d}) + (1-\theta_1-\theta_2)C(\gamma_{rd}),$$
$$(\theta_2 C(\gamma_{s_2 d}) + (1-\theta_1-\theta_2)C(\gamma_{rd}))/\sigma,$$
$$(\theta_1 C(\gamma_{s_1 d}) + \theta_2 C(\gamma_{s_2 d}) + (1-\theta_1-\theta_2)C(\gamma_{rd}))/(1+\sigma)\} \quad (4)$$

where $\sigma = R_2'/R_1'$.

Here, we consider a symmetric MARC, i.e., $\sigma = 1$, and an equal time-allocation for the two sources and the relay ($\theta_i = \theta_r = 1/3$). Also, we assume that $\gamma_{s_1 r} = \gamma_{s_2 r}$ and $\gamma_{s_1 d} = \gamma_{s_2 d}$. The achievable rates $R_1'$ and $R_2'$ are given by

$$R_1' = R_2' = \frac{1}{3}C(\gamma_{sd}) + \frac{1}{6}C(\gamma_{rd}) \quad (5)$$

A similar analysis can be performed for the multi-user scenario with $q$ sources. The achievable rates for sources $s_i$, $i = 1,\ldots,q$, with equal time-allocation ($\theta_i = \theta_r = 1/(q+1)$) are given by

$$R_i' = \frac{1}{q+1}C(\gamma_{sd}) + \frac{1}{q(q+1)}C(\gamma_{rd}) \quad \forall i \quad (6)$$

The rate of the overall system (sum rate) is $R_{\text{eff}} = q R_i'$.

In Table I we report the minimum values of $\gamma_{sd}$ such that a system rate $R_{\text{eff}} = \frac{q}{2(q+1)}$ is achieved for a fast fading channel. In the computation, we assumed $d_{sr} = (1/4)d_{sd}$, $d_{rd} = (3/4)d_{sd}$, and BPSK modulation.

## VI. EXIT CHARTS ANALYSIS

The equivalent representation of the wireless relay network in Fig. 3 allows applying tools for the analysis of SCCs to analyze the performance of the proposed distributed coding scheme. In this section, we compute the convergence thresholds of the proposed coding scheme through an extrinsic information transfer (EXIT) charts analysis [12]. For convenience, the equivalent representation of the multi-user wireless network in Fig. 3 can be further redrawn in the form of Fig. 4, where the outer encoder is not directly connected to the channel. Then, the contribution of the direct link between the sources and the destination is moved to the inner encoder $\mathcal{C}_I'$ (through its systematic branch). Note that the link $l_{sd}'$ corresponds to the concatenation of $l_{sr}$ and $l_{dr}'$ in Fig. 4. Also, note that for analysis purposes, representations in Figs. 3 and 4 are equivalent.

We assume that $\gamma_{sr}$ is high enough, so that no errors occur at the relay. In this case, the convergence behavior of the distributed turbo-like code in Fig. 2 can be tracked using standard EXIT charts by plotting in a single chart the EXIT curve of the outer encoder $\mathcal{C}_O$ (which is now independent of $\gamma_{sd}$) and the EXIT curve of the inner encoder $\mathcal{C}_I'$ which depends on both $\gamma_{sd}$ and $\gamma_{rd}$. In Fig. 5 we plot the EXIT charts of the proposed distributed coding scheme for strategy B and $q = 2$ and $q = 4$ users over a fast fading Rayleigh channel. The same 4-state, rate-1/2, recursive convolutional encoder with generator polynomials $(1, 5/7)$ in octal form is used at each source. The 4-state, rate-1, recursive convolutional encoder with generator polynomial $(5/7)$ is used at the relay. Also, we assume that $\rho = 1/q$, i.e., only one bit out of $q$ at the output of encoder $\mathcal{C}_e$ is preserved. Note that with this choice of $\rho$ the throughput of the link $l_{rd}$ is kept constant independently of the number of users, and $R_{\text{eff}} = \frac{q}{2(q+1)}$. Finally, we assume that $d_{sr} = (1/4)d_{sd}$ and $d_{rd} = (3/4)d_{sd}$, which translate into $g_{sr} = 21.19$ dB and $g_{rd} = 4.4$ dB, respectively (see Section II-A). A tunnel between the EXIT curve of the outer encoder and the EXIT curve of the inner encoder opens at $\gamma_{sd} = 0.77$ dB and $\gamma_{sd} = 1.57$ dB for $q = 2$ and $q = 4$ users, respectively, indicating convergence around these values. The estimated convergence thresholds are given in Table I for $q = 2, 4, 8, 16, 20, 30, 50$ and $100$ users and strategies A and B. For strategy A, the same 4-state, rate-1/2, feedforward convolutional encoder with generator polynomials $(5, 7)$ is used at each source, and the 4-state, rate-1, recursive convolutional encoder with generator polynomial $(3/7)$ is used for $\mathcal{C}_e$. Also, we set $J = q$, i.e., $R_{\text{eff}} = \frac{q}{2(q+1)}$. The estimated convergence thresholds are similar for both strategies. The predicted thresholds match with simulation results for very long block sizes, even if (some) errors occur at the relay (i.e. when $\gamma_{sr}$ is limited). The proposed distributed turbo-like code performs within $1.5 - 2.0$ dB from capacity for a number of users up to $q = 20$. For higher number of users the gap to capacity increases.

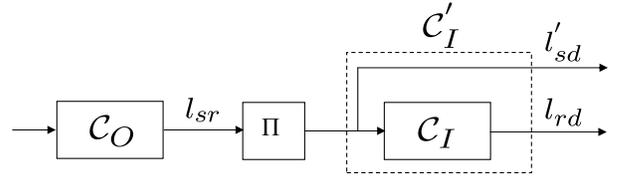

Fig. 4. Another equivalent representation of the distributed turbo-like code of Fig. 2.

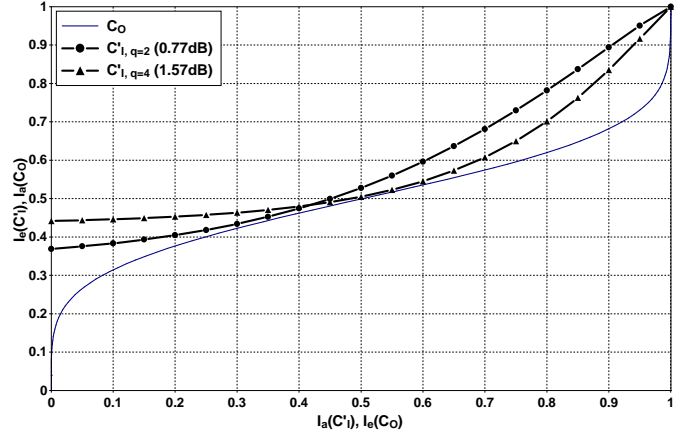

Fig. 5. EXIT chart of the proposed distributed turbo-like code, strategy B, for $q = 2$ and $q = 4$ users. Fast fading Rayleigh channel.

## VII. SIMULATION RESULTS

The performance of the proposed scheme is evaluated through simulations. For simplicity, we assume that each source uses the same encoder. We consider the same encoders as in Section VI for strategies A and B. The extension of the proposed scheme to other constituent encoders at the sources, e.g., block codes, is straightforward. We also assume $J=q$ and $\rho=1/q$ for strategy A and B, respectively, as in Section VI. Therefore, in both cases $R_{\text{eff}} = \frac{q}{2(1+q)}$.

TABLE I
CONVERGENCE THRESHOLDS FOR THE DISTRIBUTED TURBO-LIKE
CODING SCHEME AND DIFFERENT NUMBER OF USERS

|  | Threshold (A) | Threshold (B) | Capacity |
|---|---|---|---|
| $q=2$ | 0.47 dB | 0.77 dB | -1.0982 dB |
| $q=4$ | 1.37 dB | 1.57 dB | 0.0658 dB |
| $q=8$ | 2.32 dB | 2.42 dB | 0.8604 dB |
| $q=16$ | 3.27 dB | 3.27 dB | 1.3173 dB |
| $q=20$ | 3.52 dB | 3.52 dB | 1.4152 dB |
| $q=30$ | 4.00 dB | 4.00 dB | 1.5406 dB |
| $q=50$ | 4.66 dB | 4.66 dB | 1.6618 dB |
| $q=100$ | 5.25 dB | 5.40 dB | 1.7443 dB |

In Fig. 6 we give frame error rate (FER) results for strategies A and B and block length $k_i = k = 96$ bits for $q = 2, 4, 8, 16, 20, 30, 50$ and 100 users over a Rayleigh fast fading channel as a function of $\gamma_{sd}$. An S-random interleaver and a maximum of fifteen decoding iterations were considered. We also assumed that $d_{sr} = (1/4)d_{sd}$ and $d_{rd} = (3/4)d_{sd}$. While the block length $k$ is short the overall distributed turbo-like code turns out to be very powerful. Very low error rates are achieved for all values of $q$. Both strategies show similar performance. Note that the curves shift to the right with increasing number of users, since the effective code rate is higher. Note also that the curves get closer to the predicted convergence thresholds for increasing values of $q$. This result was expected, since the interleaver length (and therefore the block length of the overall SCC) increases with the number of users. Clearly, performance will break down if a very large number of users is considered.

For comparison purposes, we also consider the non-cooperation scenario where multiple sources transmit to the destination without the help of the relay. For fair comparison, we plot two curves, assuming that each source transmits with a rate 1/3 (which corresponds to the effective rate for two users) and with rate 1/2 (the effective rate tends to $R_{\text{eff}} = 1/2$ for increasing $q$). The proposed distributed turbo-like code shows a significant gain with respect to the non-cooperation scenario. Large gains were also obtained for the Gaussian and the block fading channels.

## VIII. CONCLUSIONS AND FUTURE WORK

In this paper, we proposed a distributed turbo-like coding scheme for a wireless network with relays where multiple sources transmit to a destination with the help of a relay. The relay decodes the transmitted codewords from all sources and properly combines and re-encodes them to generate some extra redundancy. The amount of redundancy transmitted by the relay can be easily adjusted according to error rate requirements and throughput and/or power constraints through a tuning parameter (parameter $J$ and parameter $\rho$ for strategy A and B, respectively). The proposed scheme can be regarded as a distributed SCC. Therefore, at the receiver a decoding strategy that resembles decoding of SCCs can be used. The proposed scheme achieves very low error rates even for a large number of users and a significant gain with respect to the non-cooperation case. Furthermore, it allows to achieve a high degree of flexibility in terms of code rate, block length and number of users. For instance, one can easily consider that each user uses a different code with a different rate and block length.

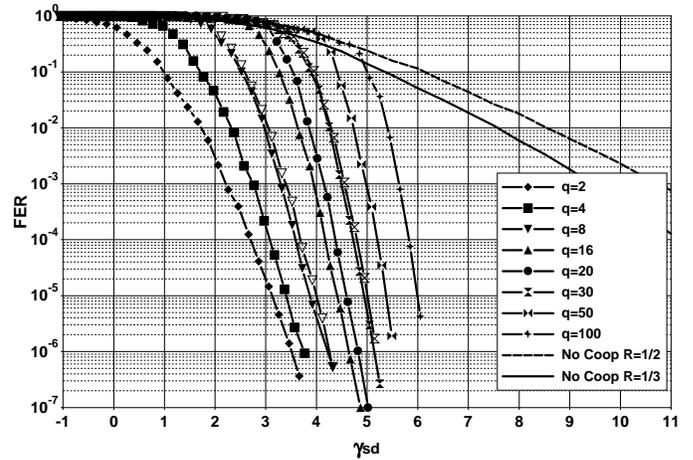

Fig. 6. FER curves for the turbo-like coding scheme of Fig. 2 using strategy A (solid markers) and B (empty markers) for $q = 2, 4, 8, 16, 20, 30, 50$ and 100 users over Rayleigh fast fading channel. $k_i = k = 96$ bits, 15 iterations.

Future work includes optimization of the time allocation (parameters $J$ and $\rho$) and the extension of the proposed scheme to a scenario with multiple-access interference.

## REFERENCES

[1] E. C. van der Meulen, "Three-terminal communication channels," *Adv. Appl. Prob.*, vol. 3, pp. 120–154, 1971.
[2] M. Valenti and B. Zhao, "Distributed turbo codes: towards the capacity of the relay channel," in *Proc. IEEE Vehicular Technology Conference (VTC)*, pp. 322–326, Oct. 2003.
[3] C. Hausl, F. Schrechenbach, and I. Oikonomidis, "Iterative network and channel decoding on a Tanner graph," in *Proc. 39th Annual Allerton Conf. on Commun., Control, and Computing*, Sept. 2005.
[4] S. Yang and R. Koetter, "Network coding over a noisy relay : a belief propagation approach," in *Proc. IEEE Int. Symp. Inf. Theory (ISIT)*, pp. 801–804, 2007.
[5] R. Pyndiah, A. Kabat, K. A. Cavalec, and F. Guilloud, "Procede de transmission d'un signal numerique entre au moins un emetteur et au moins un recepteur, mettant en oeuvre au moins un relais, produit programme et dispositif relais correspondant," April 2008.
[6] M. D. Yacoub, *Foundations of Mobile Radio Engineering*. CRC Press, 1993.
[7] T. Rappaport, *Wireless Communications*. McGraw Hill International Editions series, 1999.
[8] H. Holma and A. Toskala, *WCDMA for UMTS*. Wiley, Inc., 2001.
[9] K. M. Chugg, P. Thiennviboon, G. D. Dimou, P. Gray, and J. Melzer, "A new class of turbo-like codes with universally good performance and high-speed decoding," in *Proc. IEEE Military Communications Conference (MILCOM)*, pp. 3117–3126, Oct. 2005.
[10] A. Graell i Amat, G. Montorsi, and F. Vatta, "Design and performance analysis of a new class of rate compatible serially concatenated convolutional codes," *IEEE Trans. Commun.*, vol. 57, pp. 2280–2289, Aug. 2009.
[11] C. Hausl, *Joint Network-Channel Coding for Wireless Relay Networks*. PhD thesis, 2008.
[12] S. ten Brink, "Convergence behaviour of iteratively decoded parallel concatenated codes," *IEEE Trans. Commun.*, vol. 49, pp. 1727–1737, Oct. 2001.